\documentclass{article}
\usepackage{spconf,amsmath,graphicx}

\usepackage{hyperref}
\usepackage{url}
\usepackage{enumitem}
\usepackage{xcolor}
\usepackage{multirow}
\setlist{nosep, leftmargin=14pt}
\usepackage{float}
\usepackage{mwe}

\title{Carotid artery wall segmentation in ultrasound image sequences using a deep convolutional neural network}

\name{Nolann Lain{\'e}\,$^{\star}$ \qquad 
Guillaume Zahnd\,$^{\dagger}$ \qquad 
Herv{\'e} Liebgott\,$^{\star}$ \qquad
Maciej Orkisz\,$^{\star}$} 
\address{\small $^{\star}$ Univ Lyon, INSA-Lyon, Université Lyon 1, UJM-Saint Etienne, CNRS, Inserm, CREATIS UMR 5220, U1206, F-69621, Lyon, France\\
\small $^{\dagger}$ Institute of Biological and Medical Imaging, Helmholtz Zentrum M{\"u}nchen, Neuherberg, Germany}

\begin{document}
\maketitle
\begin{abstract}
The objective of this study is the segmentation of the intima-media complex of the common carotid artery, on longitudinal ultrasound images, to measure its thickness.
We propose a fully automatic region-based segmentation method, involving a supervised deep-learning approach based on a dilated U-net network.
It was trained and evaluated using a 5-fold cross-validation on a multicenter database composed of $2176$ images annotated by two experts. 
The resulting mean absolute difference ($<120\,\mu$m) compared to reference annotations was less than the inter-observer variability ($\approx 180\,\mu$m).
With a $98.7\%$ success rate, \emph{i.e.}, only $1.3\%$ cases requiring manual correction, the proposed method has been shown to be robust and thus may be recommended for use in clinical practice. 
\end{abstract}


\section{Introduction}
\label{sec:intro}

\begin{figure}[htb]
\begin{minipage}[b]{1.0\linewidth}
  \centering
  \centerline{\includegraphics[width=8.5cm]{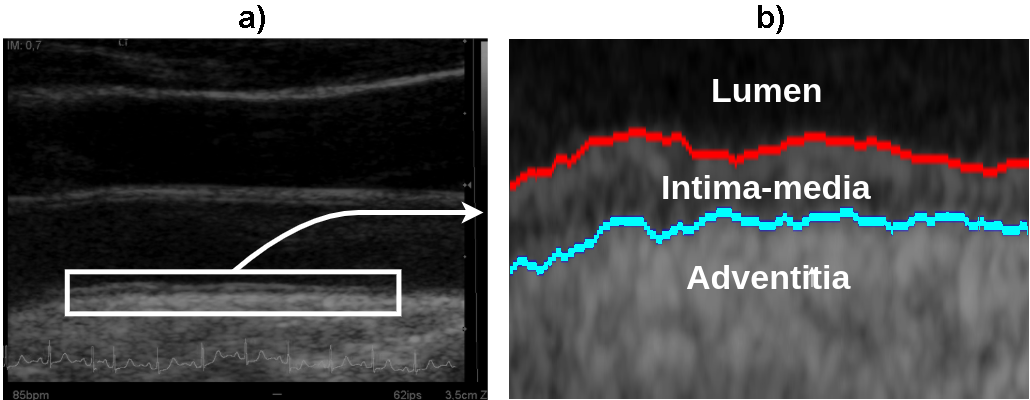}}
\end{minipage}
\caption{Example of \emph{IMC} segmentation by the proposed method: \emph{a})~Longitudinal view of the CCA in a B-mode ultrasound image. \emph{b})~Enlarged region detailing the intima–media complex with its interfaces, LI (red) and MA (cyan).}
\label{fig::IMC delineation}
\end{figure}

According to the World Health Organisation, cardiovascular diseases (\emph{CVDs}), particularly atherosclerosis, are considered the leading cause of death worldwide~\cite{kaptoge2019world} although they are preventable~\cite{mcgill2008preventing}. Prevention requires screening by means of a non-ionising and inexpensive imaging modality. Ultrasound (\emph{US}) imaging has these characteristics and is routinely used to explore the common carotid artery (\emph{CCA}), which is often considered as the sentinel of atherosclerosis~\cite{rizi2020carotid}. An early sign of this disease onset is the arterial wall thickening. To measure the thickness of interest, the contours of the intima-media complex (\emph{IMC}), namely, lumen-intima (\emph{LI}) and media-adventitia (\emph{MA}) interfaces, need to be identified (Fig.~\ref{fig::IMC delineation}).

The majority of methods in the literature use contour-based approaches \cite{delsanto2007characterization, loizou2007snakes, zahnd2017fully, raj2020automated} to exploit the intensity peaks caused by the echoes at the interfaces, named double-line pattern. Region-based approaches are less used and combine despeckling with threshold-based segmentation~\cite{nagaraj2018segmentation, wang2020fully}. 

Recently, deep-learning (\emph{DL}) has been successfully used in vascular US-image segmentation to enhance the structures of interest prior to the actual delineation by more conventional contour-based methods~\cite{menchon2016early, qian2020segmentation,shin2016automating}.
The drawback of these approaches is the necessity to combine a learnable pre-processing operation with an analytic segmentation task.

The main contribution of the present work is a supervised learnable segmentation method, designed to extract the two contours of the \emph{IMC} in B-mode \emph{US} images. 
Anatomical interfaces, in asymptomatic arteries without plaque, are localized by a region-based approach using a collection of overlapping patches. 
The proposed patch-based solution successfully addresses the challenge of segmenting with a unique network architecture the entire exploitable part of the \emph{IMC} despite strong variations in its width from one image to another.

\section{Data and method}

\begin{figure*}[t]
    \centering
    \includegraphics[width=\linewidth]{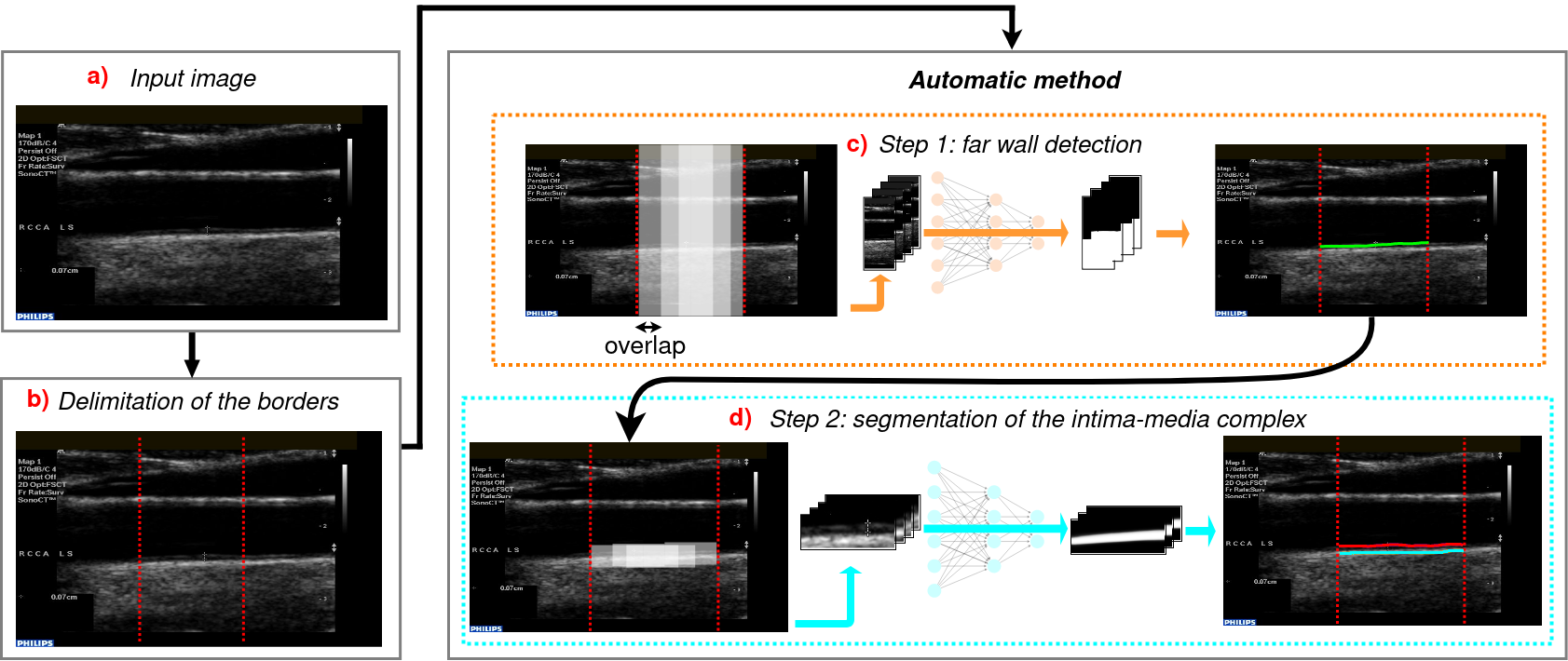}
    \caption{Flowchart of the proposed method. \emph{a})~Input image. \emph{b})~User delimitation of the left and right borders of the \emph{ROI}. \emph{c})~Far wall detection: patches are extracted through a sliding window with overlap within \emph{ROI} borders; post-processing of predicted overlapping masks leads to extraction of the median axis (green). \emph{d})~\emph{IMC} segmentation: overlapping patches are picked along the median axis and post-processing of the predicted masks leads to extraction of the \emph{LI} (red) and \emph{MA} (cyan) interfaces.}
    \label{fig::pipelines}
\end{figure*}

All training and evaluation processes were carried out on a publicly available multi-center database~(\url{http://dx.doi.org/10.17632/fpv535fss7.1}). Images were acquired from both sides of the neck, for a total of $2176$ images. Refer to~\cite{meiburger2021carotid} for more details.

For this study, we considered two experts (\emph{A1} and \emph{A2}) who independently selected a region of interest (\emph{ROI}), where interfaces were perceptible, and traced control points within it. To obtain smooth contours piecewise cubic Hermite interpolating polynomial (\emph{PCHIP}) was applied using \emph{MATLAB, Version 2020b} (The Math Works, Inc.). 

The proposed solution builds on a convolutional neural network known as U-net~\cite{ronneberger2015u}, with dilated convolutions on the bottleneck to increase its receptive field~\cite{meshram2020deep}. As the annotations are available for \emph{ROIs} of variable width, we cut the \emph{ROI} into fixed-size horizontally overlapping patches, and it is also a way to apply the same receptive field on each image with a control pixel size.  
A post-processing combines the predictions made within the patches to extract smooth contours over the entire \emph{ROI} regardless of its width.
The core of the method consists of two steps: approximately detecting the far wall (Fig.~\ref{fig::pipelines}\emph{\textcolor{red}{c}}), and precisely segmenting the \emph{IMC} contours (Fig.~\ref{fig::pipelines}\emph{\textcolor{red}{d}}). \\
Code is available at \url{https://github.com/nl3769/caroSegDeep}.

\subsection{Detection of the far wall} \label{part::Detection of the far wall} 

\begin{figure}
\centering
\begin{minipage}[b]{0.7\linewidth}
  \centering
  \includegraphics[width=\linewidth]{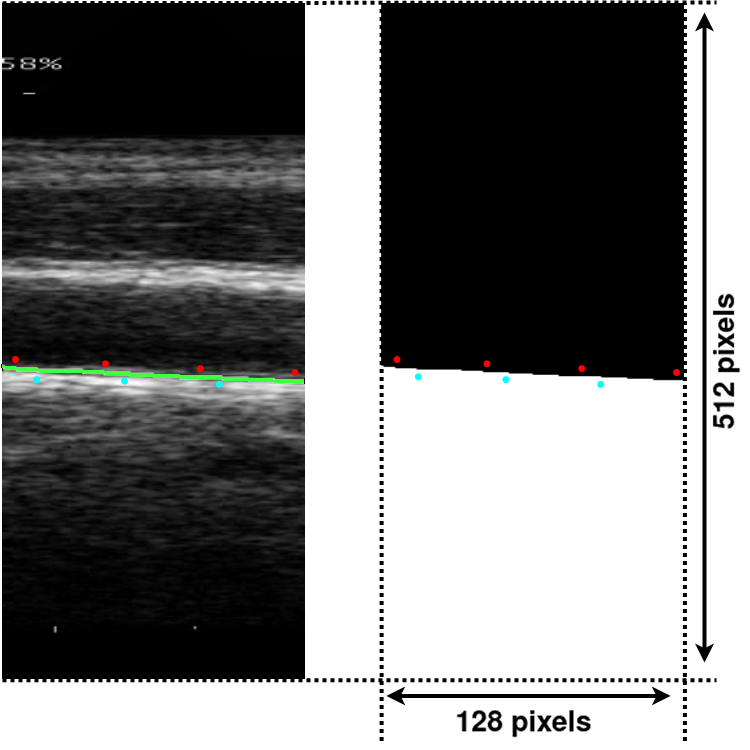}
\end{minipage}
\caption{
Data used during the far wall detection training phase.
Red and cyan dots represent the annotations for \emph{LI} and \emph{MA} interfaces, respectively. The green curve is the median axis calculated from the interpolated annotations.
}
\label{fig::Data used during the training phase (FW)}
\end{figure}

Like in many state-of-the-art methods~\cite{zahnd2017fully, wang2020fully, menchon2016early, qian2020segmentation}, the far wall is first detected and is considered as the initialization step. 
Here, the patches are of full image height and $128$-pixel width, and the corresponding U-net will be referred to as $\Theta_{FW}$.     
We first describe the pre-processing and the training phase, then we specify the post-processing chosen to obtain the curve approximately localizing the far wall on the entire \emph{ROI} width from patch-wise predictions inferred using $\Theta_{FW}$.\\ \\
\textbf{Pre-processing and training:} All images of the database were resampled to a constant height of $512$~pixels 
(as the native height of all images in the database is around $600$ pixels, the distortion thus introduced was minimal).
For training data, the median axis of the \emph{IMC} was defined as the line halfway between \emph{LI} and \emph{MA} annotations, interpolated across the entire width of the \emph{ROI}, and a reference mask ($M_{ROI}$) was generated by setting all pixels below the median axis to~$1$ and the others to~$0$. Then \emph{ROI} and  $M_{ROI}$ were identically cut into patches and a 100-pixel overlap between patches aimed at data augmentation.
Thus obtained patches with their associated masks (Fig.~\ref{fig::Data used during the training phase (FW)}) were fed into the training process, which used the ADAM optimizer and a loss function experimentally chosen to minimize the Hausdorff distance and maximize the overlay with respect to the reference masks, namely, the sum of the binary cross-entropy and the Dice loss.

\noindent \textbf{Inference and post-processing:}
Prior to inference, each image is resampled as described above, and then the corresponding \emph{ROI} is cut into $128\times 512$-pixel patches.
Next, all patches are segmented using $\Theta_{FW}$. Knowing the location and the size of each patch, two maps are created: 
\begin{itemize}
    \item \textbf{\emph{prediction map:}} contains, for each pixel, the sum of values predicted by $\Theta_{FW}$. 
    \item \textbf{\emph{overlay map:}} contains, for each pixel, the number of overlapping patches it belonged to.
\end{itemize}
\par
\noindent Dividing the prediction map by the overlay map provides, for each pixel, an average value in the range $\left[0, 1\right]$, which is then binarized by using a threshold of 0.5, to obtain the segmentation map. The latter is cleaned by retaining the largest connected component. The median axis we seek is the upper boundary of thus segmented region.  Eventually, a third order polynomial regression is applied to the retrieved boundary with the aim of increasing the robustness of the method.

\subsection{Segmentation of the IMC} \label{part::Segmentation of the IMC}

\begin{figure}
\centering
\begin{minipage}[b]{1.0\linewidth}
  \centering
  \includegraphics[width=8.5cm, trim={0 0 3.5cm 0},clip]{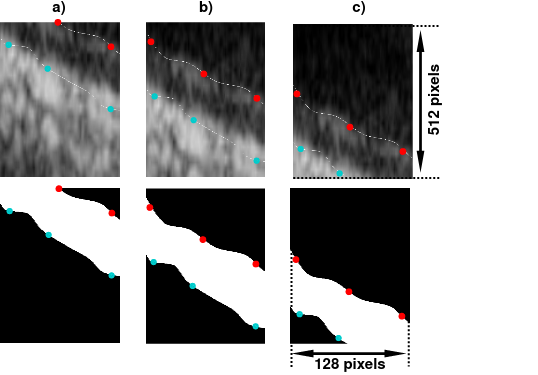}
\end{minipage}
\caption{Data used during the training phase for \emph{IMC} segmentation. Patches and their associated masks located at: \emph{a}) $\left(x_i, y_i - 128 \right)$, \emph{b}) $\left(x_i, y_i \right)$, and \emph{c}) $\left(x_i, y_i + 128 \right)$. Red and cyan dots represent the corresponding annotations for \emph{LI} and \emph{MA} interfaces, respectively. The dashed curves were obtained by interpolating the annotations.}
\label{fig::Data used during the training phase (IMC)}
\end{figure}

The far wall approximation is used to initialize the actual segmentation of the \emph{IMC}, which uses many similar concepts explained in Section~\ref{part::Detection of the far wall}: overlapping patches of $128\times 512$ pixels, an overlay map, a prediction map, a similar post-processing except that two contours are extracted (the \emph{LI} and \emph{MA} interfaces), as well as the same optimizer, loss function, and U-net architecture. The dilated U-net trained here will be referred to as $\Theta_{IMC}$. Hereafter, we emphasize the specific choices made for this step.
\\ \\
\textbf{Pre-processing and training:} The segmentation task has to be as accurate as possible, hence the algorithm works at a sub-pixel resolution. To this purpose, the vertical pixel size of the images was homogenized to $5\,\mu$m using a linear interpolation. According to this physical size, the patch height of $512$ pixels roughly corresponds $2.6\,$mm, which aims to encompass the \emph{IMC}, knowing that the average \emph{IMC} thickness is about $0.8\,$mm. For training, the ground truth was then deduced from thus interpolated images (Fig.~\ref{fig::Data used during the training phase (IMC)}): each pixel located between the annotated \emph{LI} and \emph{MA} interfaces was set to $1$, and the others to $0$. Unlike the far wall detection, the patches were extracted along the median axis: 
at each abscissa $x_i$, the mean ordinate $y_i$ of the median axis was computed on the patch width and three patches were extracted, respectively centered at $y_i$, $y_i + 128$, and $y_i-128$. This data augmentation attempted to cope with possibly inaccurate far-wall approximation as well as with tilted arteries. 
\\ \\
\textbf{Inference and post-processing:} During inference, the patches are extracted along the far-wall approximation resulting from the first step (Section~\ref{part::Detection of the far wall}). 
At each abscissa $x_i$ three or more patches are captured at different ordinates, depending on the tilt of the median axis.
The predictions made by $\Theta_{IMC}$ in all patches are combined into a prediction map, and then the segmentation map is derived thereof, as described above. 
Finally, the \emph{LI} and \emph{MA} interfaces are respectively defined as the upper and lower boundaries of thus segmented region.

\section{Results}

The evaluation was carried out using 5-fold cross-validation, so as to assess each network on data not seen during its training. In each fold, the database was split into training ($60\%$), validation ($20\%$), and testing ($20\%$) subsets. Thus, five pairs of networks $\Theta_{FW}$ and $\Theta_{IMC}$ were trained and tested independently, and the results reported here are the merging of the test sets of these five pairs, thus evaluating the method on the entire database.

In the proposed cascade approach, a failure of the first step (far wall detection) will trigger a failure of the second step (\emph{IMC} segmentation). To conduct a fair evaluation of both steps, we first quantified the success rate of the first step alone, then we quantified the accuracy of the second step by manually enforcing valid initial conditions when needed. \\  \\
\textbf{Robustness of the far wall detection:}
After visual inspection, $36$ predicted median axes ($1.3\%$ of the database) were considered as failures, \emph{i.e.} curves unusable to initialize the \emph{IMC} segmentation step.
Hence, the success rate was of $98.7\%$ and in the $36$ images with failures, the median axis was manually redrawn using a home-made graphical interface.\\  \\ 
\textbf{Accuracy of the \emph{IMC} segmentation:} 
The segmentation error was quantified by measuring the column-wise median absolute difference (\emph{MAD}) between the method output and the annotations performed by \emph{A1}, for \emph{LI}, \emph{MA}, and \emph{IMT}. These results are summarized in Table \ref{tab::results of the segmentation of the IMC}.

\begin{table}[H]
\centering
\caption{
Mean absolute difference ($\pm$ standard deviation) of segmentation results, for contour locations (\textit{LI}, \textit{MA}) and thickness quantification (\textit{IMT}).
Middle column: Method against reference annotations.
Right column: Inter-observer variability.
} 
\begin{tabular}{ccc}
Measure & Method vs. \textit{A1} $(\mu m)$ & \textit{A2} vs. \textit{A1} $(\mu m)$ \\
\hline
LI  & $ 119 \pm 124$ & $ 183 \pm 160 $ \\
MA  & $ 107 \pm 120$ & $ 177 \pm 149 $ \\
IMT & $ 161 \pm 159$ & $ 254 \pm 211 $ \\
\end{tabular}
\label{tab::results of the segmentation of the IMC}
\end{table}

\section{Discussion and conclusion}

We developed and assessed an almost-automatic (two user mouse clicks
to define the limits of the exploitable \textit{ROI}) deep-learning method to extract the contours of the intima-media complex in longitudinal B-mode ultrasound images of the carotid artery. The method first approximately localizes the far wall, and then segments the anatomical interfaces of interest. The proposed approach allows segmenting \textit{ROIs} of variable width without having to resize the images.

Robustness of the far-wall localization step is a prerequisite for overall correct segmentation. This step was successful in all but $1.3\%$ of the images, showing the robustness of the method. The actual segmentation step achieved good accuracy, with errors smaller than the inter-observer variability, both in terms of mean absolute difference in contour location (less than $120\,\mu$m vs. $\sim180\,\mu$m) and of standard deviation ($\sim120\,\mu$m vs. $\sim150\,\mu$m). These results were comparable with the best-performing state-of-the-art methods evaluated on the same database \cite{meiburger2021carotid}. Based on supervised learning, our method has the potential to increase its performance by using larger and more diverse database for training, which was proved in a recent study, where our method outperformed the existing ones \cite{meiburger2022carotid}.

The largest errors occurred in the presence of calcified plaques. As the work presented here was oriented towards asymptomatic plaque-free subjects, images with plaques were not expected. Nevertheless, we anticipate that results might be improved by enriching the database with such images appropriately annotated, and subsequently re-training the networks. This avenue desserves  investigation.

In conclusion, with a $98.7\%$ success rate and the accuracy comparable to human experts, the proposed method may be recommended for use in clinical practice.

\section{Acknowledgments}
\label{sec:acknowledgments}
This work was partly supported, via NL's doctoral grant, by the LABEX PRIMES (ANR-11-LABX-0063) of Université de Lyon, within the program "Investissements d'Avenir" (ANR-11-IDEX-0007) operated by the French National Research Agency (ANR). \\
The authors have no relevant financial or non-financial interests to disclose.

\section{Compliance with ethical standards information}

The data from human subjects used in this work were obtained and treated in line with the principles of the Declaration of Helsinki. Approval was granted by the Ethics Committees of the institutions involved in creating the multicentric database, from which these data were accessed.

\bibliographystyle{IEEEbib}
\bibliography{refs_initials.bib}

\end{document}